# Spatial characteristics of nickel-titanium shape memory alloy fabricated by continuous directed energy deposition

Shiming Gao[1], Fei Weng[1], Ojo Philip Bodunde[1], Mian Qin[1], Wei-Hsin Liao[1,3*], Ping Guo[2,*]

[1]Department of Mechanical and Automation Engineering, The Chinese University of Hong Kong, Shatin, Hong Kong, China

[2]Department of Mechanical Engineering, Northwestern University, Evanston, IL, USA

[3]Institute of Intelligent Design and Manufacturing, The Chinese University of Hong Kong, Shatin, Hong Kong, China

*Corresponding authors:　Ping Guo: ping.guo@northwestern.edu
　　　　　　　　　　　　Wei-Hsin Liao: whliao@cuhk.edu.hk

## Abstract

Additive manufacturing has been adopted to process nickel-titanium shape memory alloys due to its advantages of flexibility and minimal defects. The current layer-by-layer method is accompanied by a complex temperature history, which is not beneficial to the final characteristics of shape memory alloys. In this study, a continuous directed energy deposition method has been proposed to improve microstructure uniformity. The spatial characterization of nickel-titanium shape memory alloy fabricated by continuous directed energy deposition is investigated to study the temperature history, phase constituent, microstructure, and mechanical properties. The results indicate that the fabricated specimen has a monotonic temperature history, relatively uniform phase distribution and microstructure morphology, as well as high compressive strength (2982 MPa~3105 MPa) and strain (37.7%~41.1%). The reported method is expected to lay the foundation for spatial control during the printing of functional structures.

## 1. Introduction

Nickel-titanium (NiTi), also referred to as Nitinol, has unique properties of shape memory effect, superelasticity behavior, biomechanical compatibility, and corrosion resistance [1-3]. These properties have enabled its vast application in different engineering and medical fields such as actuators, sensors, orthodontic wires, bone plates, etc. [4-7]. The fundamental mechanism for NiTi shape memory effect is the thermal or stress-induced phase transformation between the "austenite" high-temperature phase and "martensite" low-temperature phase, which is sensitive to element composition.

Currently, the widely adopted methods for Nitinol production are vacuum arc melting and vacuum induction melting, followed by casting, hot or cold working with annealing, and shape memory treatment [8]. However, vacuum arc melting requires multiple remelting steps to ensure material homogeneity, while vacuum induction melting is prone to crucible contamination. The following casting would also lead to chemical segregation defects [9]. The conventional machining methods, such as mechanical cutting, usually suffer from chip formation, cyclic hardening, and acceleration of tool wear [10,11]. Powder metallurgy methods, such as hot isostatic pressing, metal injection molding, self-propagating high-temperature synthesis, have been developed to overcome the above-mentioned problems of manufacturing Nitinol. Hot isostatic pressing was adopted to produce semi-finished components with both pre-alloyed

NiTi powders and elemental Ni and Ti powders. Unfortunately, the component's geometry was simple due to the requirement of equal pressure in all directions [12,13]. The metal injection molding method mixes binder and metal powder to obtain the feedstock for green body fabrication, which will be sintered to get the final components. The used binder, as an impurity source, is detrimental to the final properties of Nitinol [14]. Self-propagating high-temperature synthesis is concerned with the self-sustaining chemical reaction through the ignition of a compressed powder mixture. Once the reaction begins, it becomes difficult to control. Besides, components fabricated through this method are accompanied by some porosities [15]. Thus, based on the above review, conventional manufacturing technologies still have some critical limitations in the manufacture of Nitinol.

Additive manufacturing (AM) technologies, such as selective laser melting (SLM), selective electron beam melting (SEBM), and directed energy deposition (DED), have recently been investigated to fabricate NiTi parts due to their low or minimum defects and design flexibility with complex geometry [16]. Corresponding studies indicated that the process parameters [17-19], chamber oxygen level [20,21], and post heat treatment [22] have a huge influence on the microstructure evolution (e.g., crystalline shape, arrangement, preferred crystallographic orientation, and secondary phase precipitation), as well as thermal and mechanical properties (e.g., reversible martensitic transformation temperature, superelasticity, compressive/tensile strain-stress).

However, previous research on NiTi alloys rarely involved the spatial characteristics of the deposited structure, especially those manufactured by the DED method. Hamilton et al. [23] compared the microstructure and compressive shape memory effect of NiTi shape memory alloys manufactured by SLM and DED methods. The results showed that the spatial variation of the composition and grain structure in the DED specimen is greater than that of the SLM samples. The oriented/detwinned martensite in the DED sample is more unstable since its residual martensite could immediately recover upon heating, while an additional thermal input was required before the initiation of reversion in SLM samples. Bimber et al. [24] investigated the microstructure spatial distribution along the build height direction. It was found that $Ni_4Ti_3$ precipitation presented coarser morphology near the substrate, while a finer structure was observed close to the top. The digital image correlation analysis revealed that the largest transformation strain levels and diffuse strain contours occurred in the area with finer $Ni_4Ti_3$ precipitation. Hamilton et al. [25] fabricated a large volume specimen via the DED process from an elemental powder blend to investigate the spatial variation of the thermal-induced martensitic transformation temperatures. The differential scanning calorimetry (DSC) analysis results showed that the specimens ($Ti_{52.1}Ni_{47.9}$ at.%) taken from a higher build location had equivalent transformation temperatures and enthalpy measurements along the scan direction, while the sample extracted from a lower height position had a larger difference for both measurements. The above researches have shown that there are spatial variations in thermal and mechanical properties in the as-deposited samples, which usually result in an uncontrollable or multistep response of shape memory effect and superelasticity. These performance variations are likely to be associated with the layer-by-layer deposition manner, which exposes the previously solidified layers to complex thermal histories due to remelting.

The continuous DED method, as shown in Fig. 1(b) [26-28], provides a new alternative to fabricate the NiTi alloy samples. Remelting of previous layers or adjacent tracks will be avoided during the manufacturing process, which makes the thermal history monotonous. Therefore, the spatial characteristic of as-deposited NiTi alloy via continuous DED may be different from the previously observed results. Inspired by these differences in raw materials and process methods, the NiTi rods via pre-alloyed powders are fabricated by the continuous DED method to investigate their spatial characteristics in this paper. The temperature histories at the bottom, middle, and top regions are first predicted by a finite-element (FE) model with COMSOL Multiphysics software, followed by microscopic inspection to investigate the phase constituent and microstructure variation. Then, the corresponding thermal-induced martensitic transformation temperatures are presented and discussed. Finally, the mechanical properties of compressive stress-strain behavior and microhardness are evaluated. The results indicate that continuous DED can achieve a relatively homogenous spatial distribution of phase constituent, thermal responses, and mechanical properties.

## 2. Materials and methods

$Ni_{50.93}Ti_{49.07}$ at.% pre-alloyed powders with spherical morphology were used to fabricate the samples. The powder size was in the range of 20-50 μm and its chemical composition is shown in Table 1. NiTi alloy plates with dimensions of 50 mm x 50 mm x 3 mm were used as the substrate that was polished and cleaned before deposition. The specimens were fabricated in an Argon atmosphere ($O_2$ < 20 ppm) via our self-developed DED system [29].

In the traditional DED method, a single track or thin layer is firstly created in the X-Y plane, then a subsequent track or layer is deposited on the previous one accompanied by remelting and reheating. The structure then is fabricated through this repeated layer-by-layer deposition as shown in Fig. 1(a). In the continuous DED method as shown in Fig. 1(b), the scan direction is following the direction of material growth and the growth rate matches the scan speed of the laser. The scan speed and deposition rate have to be carefully matched to ensure the proper focus of the laser beam and the powder stream.

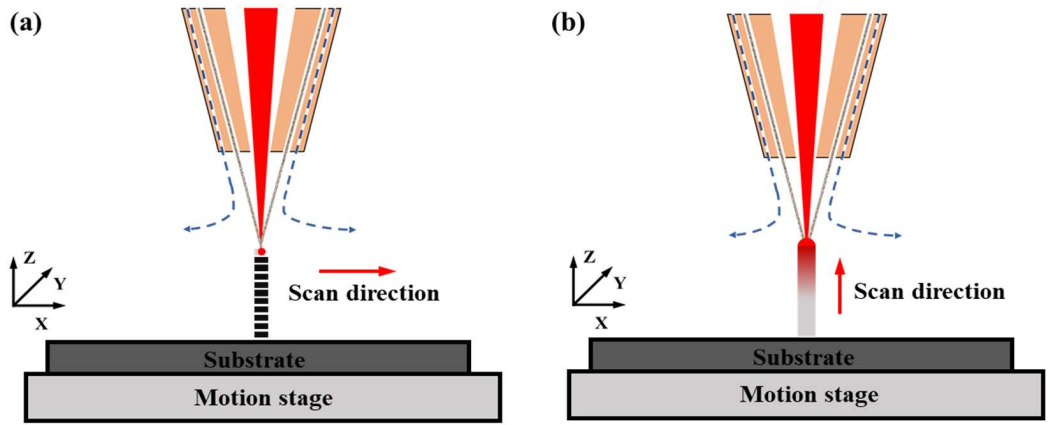

Fig. 1 Schematics of the (a) traditional layer-by-layer; and (b) continuous DED processes.

Table 1 The nominal composition of the pre-alloyed NiTi powder (wt.%).

| Alloy | Ni | Fe | C | Ti | O | N |
| --- | --- | --- | --- | --- | --- | --- |
| NiTi powder | 54.5 | 0.0074 | 0.006 | 45.4241 | 0.0607 | 0.0018 |

A series of rod samples with the designed length of 15 mm were fabricated by this continuous DED method to investigate their spatial characteristics. The deposition parameters were optimized through previous experiments and chosen as follows: laser power of 34.3 W, a Z-scan speed of 0.5 mm/s, and a powder feeding rate of 0.015 g/s. The focal spot radii for the laser and powder flux were 0.2 mm and 0.6 mm, respectively. The photos of as-deposited NiTi rods with a diameter of 1.65 mm and a length of 14.5 mm are shown in Fig. 2(a). Then three specimens from the bottom, middle, and top regions of the as-printed sample were extracted along the build direction as shown in Fig.

2(b). For each specimen, the following thermal and mechanical analyses were conducted.

The phase constituents were analyzed by an X-ray diffractometer (Smartlab, *Rigaku, Japan*). The microstructure was investigated by a high-resolution optical microscope (RH-2000, *Hirox, Japan*) and a scanning electron microscope (JSM 6400, *JEOL, Japan*). The scanning electron microscope with Energy-dispersive X-ray spectroscopy (EDS) detector was also used to capture the chemical composition of a specific point or area. The samples for microstructure observation were first mechanically polished through SiC sandpaper (180, 400, 600, 800, and 1200 grit), and then polished using 0.5 μm diamond suspensions on synthetic cloths. The etchant was a volume mixed solvent of HF (2%), $HNO_3$ (10%), and $H_2O$ (88%), respectively. The transformation temperatures (TTs) were detected by differential scanning calorimetry (DSC 3, *Mettler Toledo, Switzerland*) with heating and cooling rates of 10°C/min from -80°C to 80°C. The compressive mechanical properties of fabricated samples were evaluated by uniaxial monotonic compression tests at a rate of 0.2 mm/min on a uniaxial compressive machine (Alliance RT/50, *MTS, America*) at room temperature. The oil was utilized to lubricate the bottom and top surfaces to reduce the radial friction. For repeatability, each sample was measured three times for the compression tests. The microhardness was investigated using a Vickers indenter (IndentaMet 1114, *Buehler, USA*) with a load of 100 g and a dwell time of 15 s. The measurement points were evenly distributed with a 1 mm pitch along the scan direction. Three indents were taken for each point of the sample analyzed.

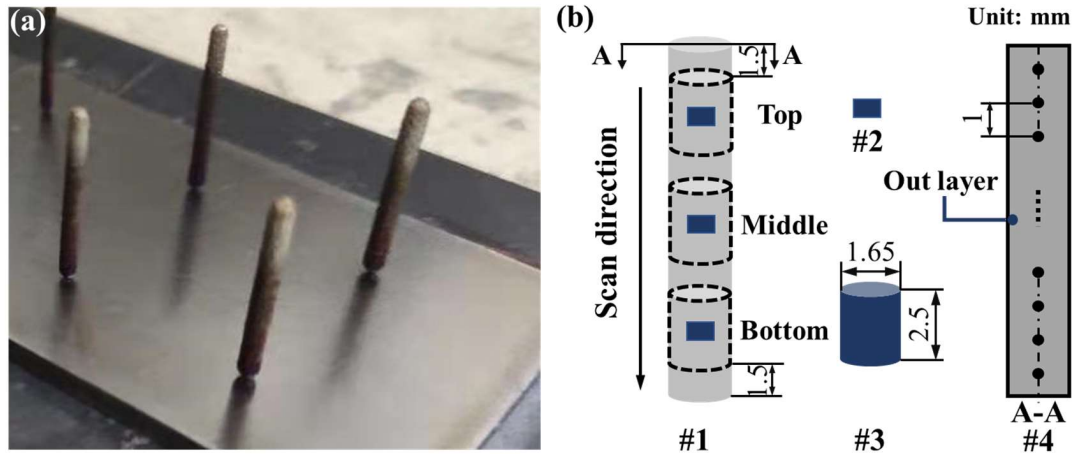

Fig. 2 (a) Photo of the as-deposited NiTi rod structures; and (b) schematic illustration of the as-deposited specimens for thermal and mechanical testing: (#1) as-deposited, (#2) DSC, (#3) compression, and (#4) microhardness.

A 2D finite-element (FE) model with COMSOL Multiphysics 5.4 was developed to investigate the temperature history of the continuous DED process. The proposed heat transfer model accounted for several physical phenomena in the complex continuous DED processes including heat conduction, convection, radiation, and phase transition. The substrate and powders were chosen as pre-alloyed NiTi ($Ni_{50.93}Ti_{49.07}$ at.%). The laser beam was assumed to have a Gaussian distribution. As shown in Appendix, the temperature-dependent material properties were adopted for the simulation, which was important for the accurate prediction of thermal history. The power effective absorption coefficients were determined through experiments.

## 3. Results and discussions

The continuous DED method avoids the remelting of the previous layers or adjacent tracks by eliminating the intermittent steps. It is assumed that the inter-layer defects, such as inhomogeneity around the molten pool boundary, inter-layer porosity, and grain dislocation, will be alleviated. Compared to the traditional layer-by-layer method, the temperature history will become monotonous. The microstructure, phase constituent, and thermal and mechanical properties are expected to have better uniformity. The corresponding investigations on the thermal and mechanical properties then are conducted to test these hypotheses in the following parts.

### 3.1 Temperature history

During the deposition, the NiTi rod grows at a rate equal to the scan speed while the laser source moves from the previous solidified area, causing further heat flux input to the subsequent deposition location. Due to the nature of continuous DED, commercial temperature measurement equipment such as infrared cameras and pyrometers cannot easily obtain temperature information inside the fabricated rod structure. Hence, a 2D finite-element (FE) model was developed in COMSOL Multiphysics software to investigate the location-dependent thermal history along the scan direction as shown in Fig. 3(a). The material addition was controlled by the element birth and death method with a mesh size of 20 μm. The whole deposition process is governed by energy conservation [30,31]:

$$\rho c_p \frac{\partial (T)}{\partial t} - \nabla \cdot (k \nabla T) = \dot{Q}(x, y, z) \tag{1}$$

where $\rho$ is the material density; $\dot{Q}(x, y, z)$ is the external energy source introduced by the laser power input; $c_p$ and $k$ are the heat capacity and thermal conductivity, respectively.

The simulation model was first validated by comparing the simulated melt pool sizes with the experimentally measured ones under several deposition parameters. Because the comparison between the two was difficult in the continuous deposition process, the comparison was firstly made in the traditional layer-by-layer process. The deviations of simulated and experimental results were presented in Table 2. The point 'D' located under the substrate and directly under the printed rod structure was further investigated. The temperature history of point 'D' was measured in the experiments with embedded thermocouples and compared with the simulated results. The details for the temperature measurement setup are described in the Appendix. As shown in Fig. 3(b), the two curves show a consistent trend. A maximum error of 7.7% indicates the model has acceptable credibility. The deviation mainly comes from the thermocouple response. With the validated simulation model, the three different points from the top, middle, and bottom central regions were then extracted and compared to investigate the spatial variation of the structure temperature. In the simulation process, the laser power spot keeps moving upwards at a constant speed until the end of the deposition. When the laser power spot passes through the death element, it is activated and transformed into the birth element, imitating the material growth. Since only the birth elements will be considered in the calculation process, the simulated temperature histories of the three points start at different time instances in Fig. 3(c). The X-axis represents the total manufacturing time, where the first 30 seconds are the deposition time, and the last 60 seconds are the cooling time. As shown in Fig. 3(c), it is observed that the top point has the highest temperature, but the shortest duration over 1000 K. The highest temperature is caused by heat accumulation from the previous material deposition, while the shortest duration above 1000 K is related to the least material addition and power input. The bottom point experiences the lowest temperature profile but with the longest holding time above 1000K. This phenomenon is ascribed to the colder substrate and continuous material deposition. The secondary phases such as $Ti_2Ni$ and $Ni_4Ti_3$ are prone to form when the temperature drops into the precipitation range (573 K~973 K). The deposition process itself is a self-heating treatment due to the characteristics of rapid melting and cooling rates. The formed phase composition is dependent upon the aging temperature and dwelling time. These three points from the top, middle, and bottom positions undergo a similar experience in the precipitation range, which helps to maintain similar microstructures at different locations.

**Table 2**

Simulation and experimental results at a scan speed of 2 mm/s and a powder feed rate of 0.015g/s.

| Parameters | Width (μm) | | | Height (μm) | | |
| --- | --- | --- | --- | --- | --- | --- |
| Power (W) | Sim | Exp | Dev(%) | Sim | Exp | Dev(%) |
| 34.3 | 298 | 281.5 | **5.8** | 90 | 85.6 | **5.1** |
| 45 | 398 | 380.2 | **4.7** | 129 | 118.9 | **8.4** |
| 56 | 492 | 468.0 | **5.1** | 157 | 145.4 | **8.0** |

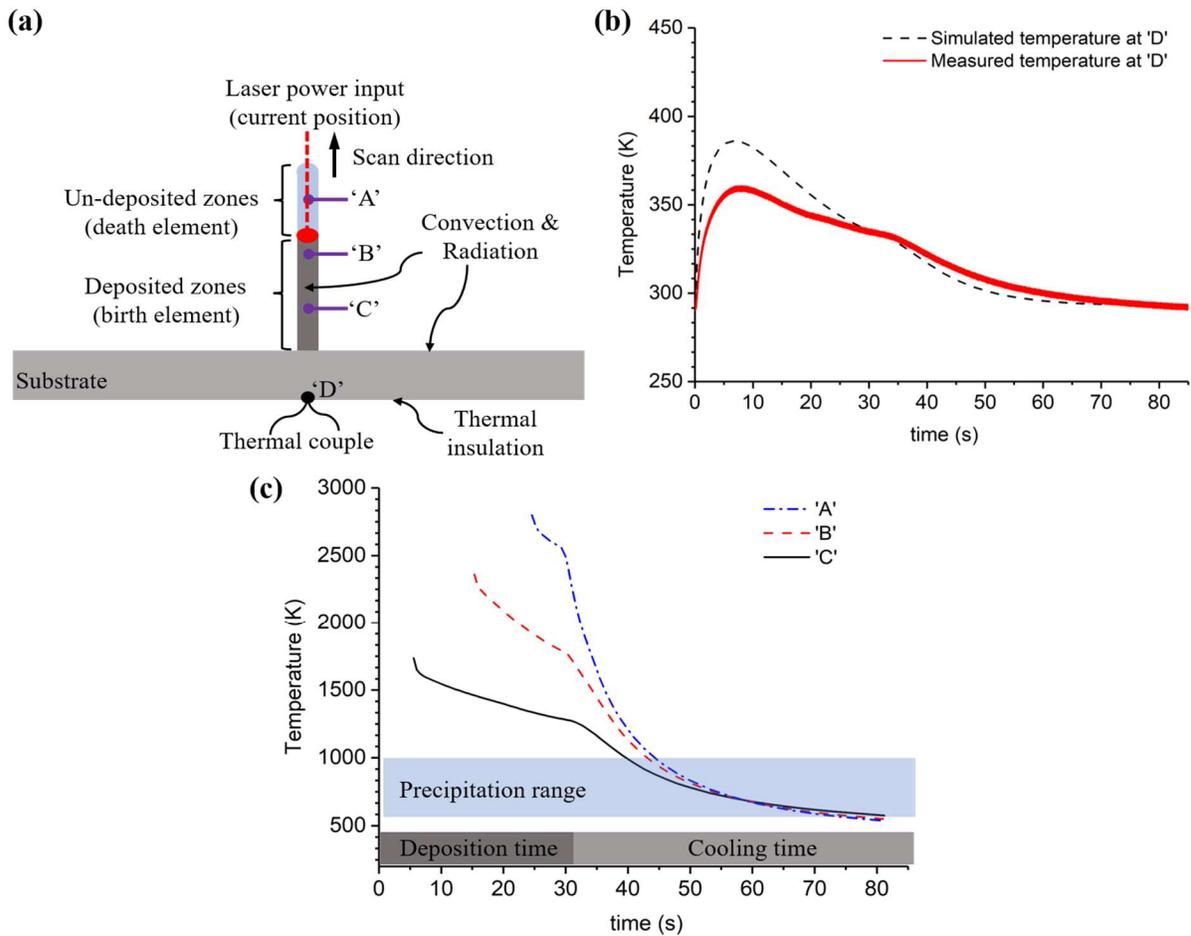

Fig. 3 (a) The schematic of simulation model; (b) measured and simulated temperature history at bottom point 'D' in (a); and (b) predicted temperature histories at the three points in (a).

## 3.2 Phase constituent and microstructure

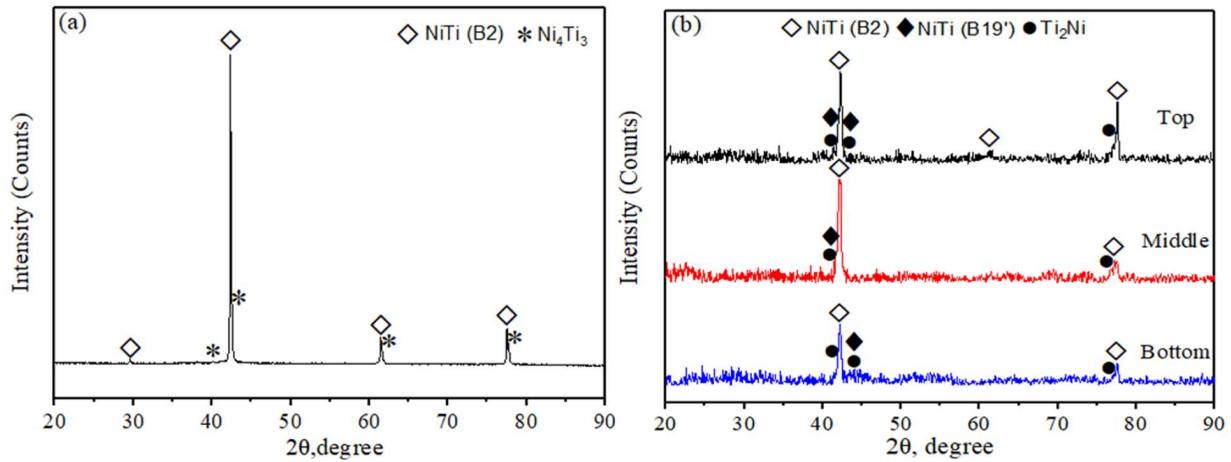

Fig. 4 XRD patterns of: (a) initial NiTi pre-alloyed powder; and (b) the DED rod specimens from top, middle and bottom regions.

Fig. 4 shows the XRD analysis results of the initial pre-alloyed powders and the NiTi alloy samples fabricated by continuous DED. The raw powders are mainly comprised of NiTi (B2) austenite phase with a tiny secondary phase $Ni_4Ti_3$ at room temperature. The existence of $Ni_4Ti_3$ particles may come from the metastable phase precipitation during powder preparation, which is usually observed for Ni-rich NiTi alloy [32]. The phase analysis of extracted DED specimens from three different locations reveals the existence of austenite phase NiTi (B2), martensite phase NiTi (B19'), and precipitation phase $Ti_2Ni$ after deposition. The corresponding three XRD results have no obvious difference between each other. The accompanied phases of NiTi (B19') and $Ti_2Ni$ are caused by the high temperature and rapid cooling rate during the deposition process. According to the binary equilibrium of the Ti-Ni phase diagram [33], a complete austenite structure should be formed at room temperature for the sample fabricated from equiatomic pre-alloyed NiTi powders. However, the quantity of Ni falling into the molten pool may be less than the initial powder ratio due to the Ni evaporation caused by powder burning when it passes through the laser beam flux [34]. This burning loss as well as the segregation of chemical composition result in the local Ti-rich liquid phase in the molten pool [35-37]. When the temperature cools down to 1310°C, the liquid NiTi alloy begins to solidify and the primary reaction occurs as shown in Fig. 5(a): liquid ($L$)→NiTi+$L$(Ti) [38,39]. The solidified NiTi phase grows rapidly in the form of columnar or equiaxed grains. When the temperature of the molten pool is further reduced to 984°C, the Ti-rich liquid phase around solidified NiTi phase then undergoes a NiTi + $L$(Ti)→$Ti_2Ni$ peritectic reaction and forms $Ti_2Ni$ precipitation around NiTi grains as shown in Fig. 5(b). As the temperature of the molten pool decreases further, a small amount of $Ti_2Ni$ secondary phase precipitates in the solid NiTi phase as illustrated in Fig. 5(c). Finally, microstructures primarily composed of a B2 austenite matrix with some irregular $Ti_2Ni$ phases in the grains and on the boundaries are formed as shown in Fig. 5(d). The martensite phase NiTi (B19') detected at room temperature is mainly related to the precipitation of $Ti_2Ni$ in the matrix phase, which impedes the propagation of reversible martensite transformation.

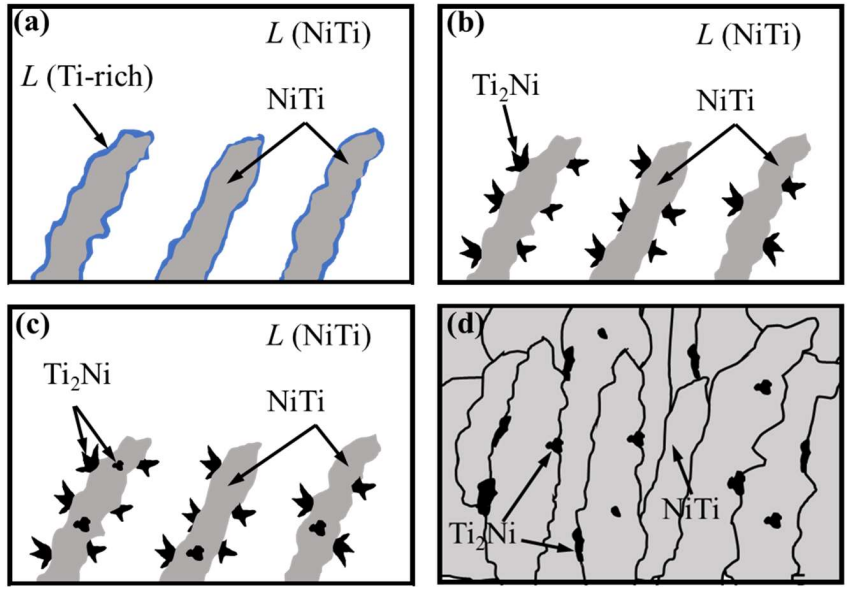

Fig. 5 Schematic illustration of solidification process of as-deposited NiTi alloy.

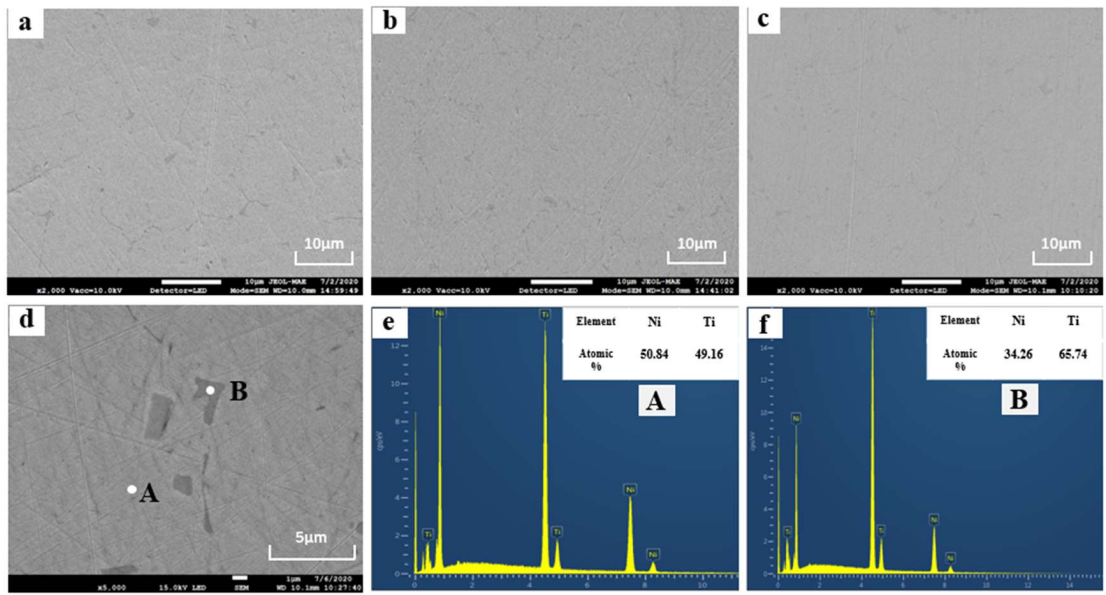

Fig. 6 Microstructure and composition analysis in the longitudinal section of as-deposited samples: (a) top, (b) middle, and (c) bottom sections; (d) magnified view of bottom section; (e) and (f) EDS analyses of the point A (light phase) and B (dark phase).

The SEM and EDS analyses were performed to further verify the XRD results. Fig. 6(a)-(c) shows the SEM images of microstructures at three different locations along the build direction of the as-deposited samples. Two different phases can be identified in all three images from the top, middle, and bottom sections: the light and dark gray phases. The light gray phases occupy the primary area, while some irregular dark gray phases are randomly distributed on the grain boundaries, which is consistent with the previous report [40]. The distribution map and area fraction of light and dark gray phases for three different regions are similar to each other. Quantitative point analysis with EDS was performed to analyze the specific element composition in the previous two-phase regions as marked in Fig. 6(d). The atomic concentration of Ni and Ti is nearly equivalent in the light gray region, which indicates the primary phase is NiTi. The ratio of Ni to Ti in the dark gray regions is close to 2:1, which indicates that the dark gray phase is $Ti_2Ni$ as identified in the XRD analysis. The martensite phase (B19') generally has a needle-like or plate-like shape [41,42]. A similar structure is also observed in our deposited sample, which is randomly distributed in the whole region due to local inhomogeneity in the material composition or stress field. The magnified SEM image is shown in Fig. 7(a) and its corresponding element distribution and quantitative point analysis are presented in Fig. 7(b)-(d). The equiatomic composition of Ni and Ti in both the two phases proves the needle-like shaped structure is the martensite phase (B19') [43,44].

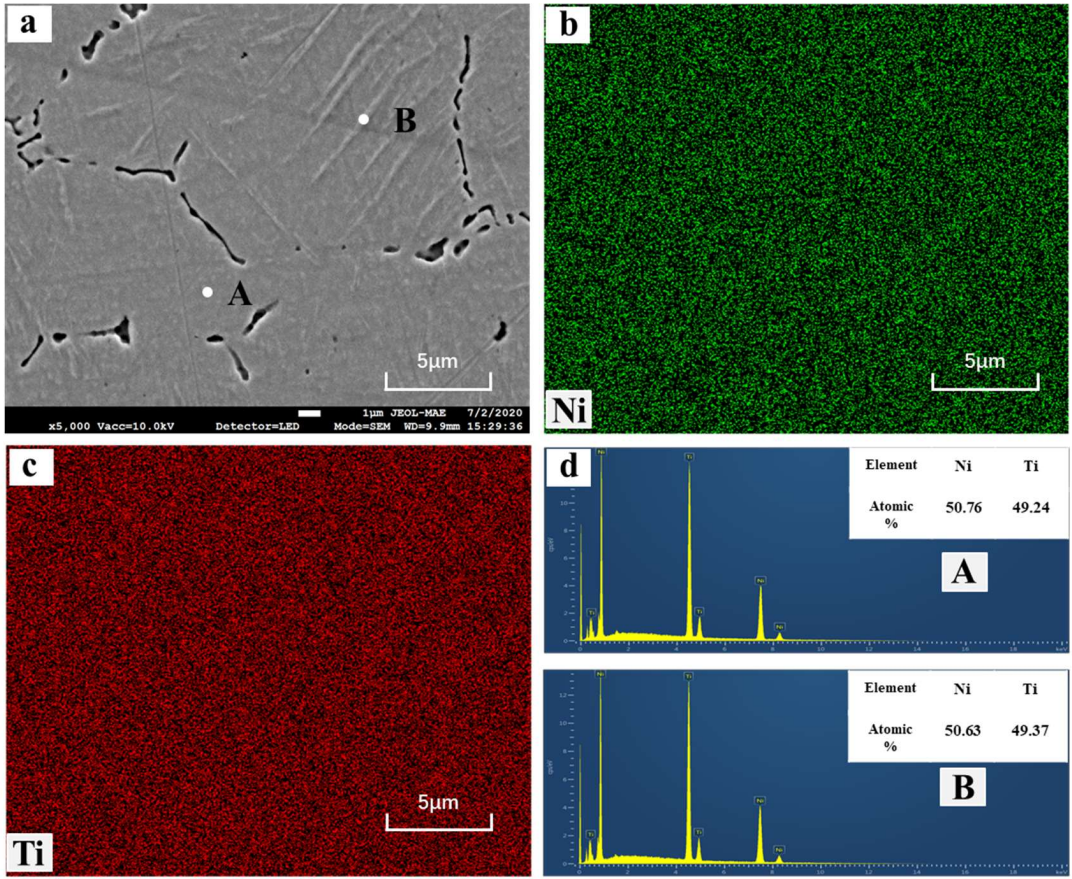

Fig. 7 Microstructure and composition analysis of the deposited sample after corrosion: (a) magnified SEM image; (b) Ni element distribution; (c) Ti element distribution; and (d) quantitative point analysis of A and B.

Fig. 8 shows the microstructures of different longitudinal cross-sections along the scan direction. The microstructure displays nearly similar grain morphologies at the three regions as shown in Fig. 8(a)-(c), while the molten pool boundaries presented in the traditional DED samples disappear. This phenomenon is attributed to the monotonous temperature history and the elimination of intermittent steps in the continuous DED process. Due to the absence of remelting from subsequent deposition as well as the consistent heat flux direction with the solidification direction, a relatively isotropic thermal conduction is obtained inside the structure during solidification. Therefore, continuous and similar equiaxed grains are formed. The morphology plotted in Fig. 8(d) shows coarse equiaxed grains at the outer layer region and a transition zone from the outer layer to the internal region. The diverse microstructure is related to the laser beam Gaussian distribution, which will result in a lower peak temperature at the outer layer than that in the internal region. Moreover, the outer layer has a faster cooling rate due to the protective gas blowing across the solidified surface. Therefore, regular austenite grains impeding secondary phase precipitation are obtained at the outer layer region. The dark discrete phase in this area is the $Ti_2Ni$ phase, which often occurs near grain boundaries [45]. Besides, there are also some defects such as voids and pores that appear in the deposited samples as shown in Fig. 8(a)-(c). The dark irregular voids are often the impurity phase that is caused by burning powder or chemical segregation during the continuous DED process. The spheroidal pores are formed due to entrapped gas bubbles in the molten pool. In comparison among these three regions, the bottom area seems to show more pores. The colder substrate causes the molten pool in the bottom area to solidify faster, resulting in less time for gas bubbles to escape. The top region seems to present larger voids. This phenomenon may be related to the gradually accumulated heat, which will change the powder absorption efficiency. The slight defocus then occurs between the laser beam and sample surface, causing ingredient fluctuation. Due to the moderate solidification time and temperature, there are fewer pores and voids in the middle region. The defects in the three regions are close to each other, which means the structure is relatively uniform.

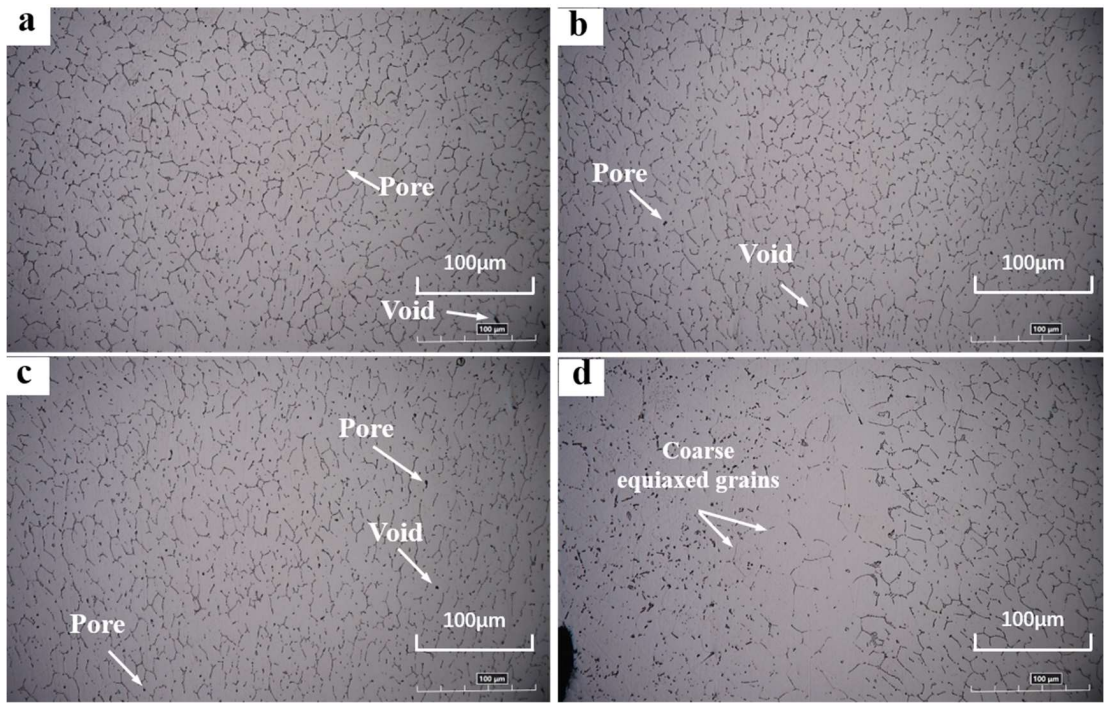

Fig. 8 Optical micrographs of longitudinal cross-sections: (a) top region; (b) middle region; (c) bottom region; and (d) outer layer.

### 3.3 Phase transformation

Fig. 9 shows the differential scanning calorimetry (DSC) curves of initial pre-alloyed NiTi powders and fabricated specimens from the top, middle, and bottom regions. The phase transformation temperatures (TTs) curves for the initial powders appear two distinct exothermic peaks in the heating and cooling cycles, which indicates a two-step transformation (B2↔R↔B19'). This phenomenon is attributed to the existence of the $Ni_4Ti_3$ phase in the NiTi matrix that has been verified by the XRD pattern in Fig. 4(a). The existence of $Ni_4Ti_3$ precipitation on grain boundaries will result in crystallographic defects such as dislocation or local stress field. These crystallographic defects will suppress the B2↔B19' transformation but promote the B2↔R transformation [46,47].

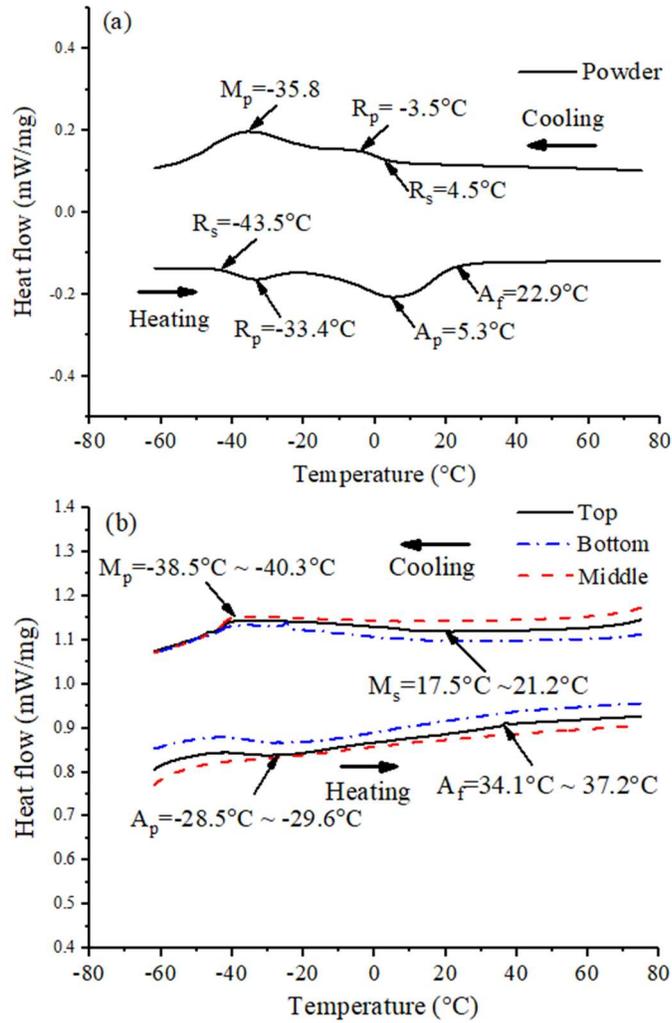

Fig. 9 DSC curves of: (a) initial powders; and (b) DED fabricated samples.

The DSC curves of fabricated specimens from the top, middle, and bottom regions all show a broad TTs range without R phase transformation. The wide TTs range indicates the gradual and suppressed phase transformation behavior that involves complex physical or chemical mechanisms. Influenced by the high laser power energy input, the heating zone will experience a rapid melting and solidification process, which results in internal residual stress in the structure. It is reasonable to assume that the internal residual stress is one of the possible reasons for the formation of defects such as lattice dislocation and lattice distortion in the as-deposited NiTi alloys [48]. Besides, chemical segregation often occurs during the DED process. The Marangoni flow in the molten pool might push Ni solute towards the edge or bottom of the pool, causing inhomogeneous chemical distribution [30,49]. The falling powder with burning loss of Ni element results in local Ti-rich liquid in the molten pool that also induces inhomogeneous composition. When the molten pool begins to solidify, this chemical segregation is prone to form secondary phases such as $Ti_2Ni$ and $Ni_4Ti_3$ in the B2 matrix and on its grain boundaries, thereby generating a local stress field around it.

The crystallographic defects then are formed [50]. These defects, especially lattice dislocations and distortions, as a barrier, will impede the martensite nucleation and interface movement [51,52]. Additional under-cooling is required to overcome this energy obstacle, resulting in a broad phase transformation temperatures (TTs) range. Compared with the DSC curve of initial pre-alloyed powders, the austenite finish temperature ($A_f$) on the heating curves is hard to be observed for all three different regions. It indicates that the martensite phase transformation to the austenite phase is still not finished at room temperature. The defects play an important role in TTs behaviors.

As shown in Fig. 9(b), the phase transformation temperatures of the three different regions all move to a lower temperature range. This decrease in TTs is probably attributed to several factors explained as follows. The possible main reason is the presence of $Ti_2Ni$ precipitations. The secondary phase of $Ti_2Ni$ will consume the Ti element from the NiTi matrix, inducing a Ni-rich matrix. The transformation temperature of fabricated NiTi alloy is very sensitive to its composition and an increase of 0.1 at % Ni will decrease 10°C transformation temperature [33]. Besides, some impurities such as oxygen and carbon might be picked up during continuous DED. Finally, the formation of previously mentioned precipitation may also cause barriers in grains or on grain boundaries that impede the propagation of reversible phase transformation front and decrease the phase transformation temperatures. The phase TTs from different locations exhibits similar behavior as shown in Fig. 9(b), which indicates the precipitation and defects of the as-deposited rod is relatively uniform.

### 3.4 Mechanical properties

Fig. 10(a) exhibits the room temperature compression behavior of as-deposited specimens from the bottom, middle, and top regions. It is interesting to notice that the specimens from different regions all show a high level of compressive strength (2982 MPa~3105 MPa) and fracture strain (37.7%~41.1%), which is higher than the previously reported results (e.g., ~2796.57 MPa and ~27.8% [53], ~2500 MPa and ~32% [54]). It is believed that the high compressive strength and fracture strain are related to the consistency of subgrain direction and compression direction [49]. Compared with the stress-strain curves of the bottom and top regions, the curve from the middle region presents a higher compressive strength before its compression fracture and a lower fracture strain when the structure fractures. This phenomenon may be related to the increased $Ti_2Ni$ precipitation in the middle region, which has not been predicted by temperature history simulation due to the formation of precipitation always involving complex physics such as inhomogeneity of chemical composition. The $Ti_2Ni$ has been reported to have a strengthening effect on the compression strength as well as a negative effect on ductility [55,56]. The corresponding microhardness at different locations on the transverse section along the scan direction is shown in Fig. 10(b). It can be seen that the microhardness in the middle region has indeed higher values than the microhardness in the bottom and top regions. The precipitation has a certain contribution to the microhardness improvement. Besides, the high and consistent compressive strength implies that the fabricated rod structure has uniform quality without major defects.

For the as-deposited NiTi samples, the compressive stress-strain curve at room temperature presents four distinguishable stages, which are marked in Fig. 10(a). Stage I indicates the elastic deformation of the austenite B2 phase. Along with the increase of compressive stress, the stress-induced austenite to martensite transformation occurs as shown in stage II. The slope of the stress-strain curve in this interval is related to the number of activated correspondence variant pairs (CPVs) [54]. Meanwhile, the elastic deformation of the transformed martensite phase and some lattice slip of the residual austenite phase may occur due to the crystal orientation and the secondary phase precipitation in the NiTi matrix. With further loading, the stress-induced austenite to martensite transformation is completed and the elastic deformation of the martensite phase happens in stage III. It has been reported that stages II and III may well overlap due to the inhomogeneous and localized stress-induced martensite transformation, caused by inhomogeneous chemical composition or impurity precipitation [53]. The plastic deformation of the martensite phase occurs in stage IV. The curve in this interval presents a normal working hardening effect with some stress fluctuation. This stress fluctuation indicates the Portevin-Le Chatelier (PLC) effect, which is mainly caused by the dislocation movement [57]. As reported, the solute atom can impede the dislocation slip to a certain extent through the pinning effect. However, when the applied stress is high enough to overcome the energy obstacle, the dislocation will pass through solute atoms and make the stress suddenly drop. This fluctuation situation continues until the sample's final fracture.

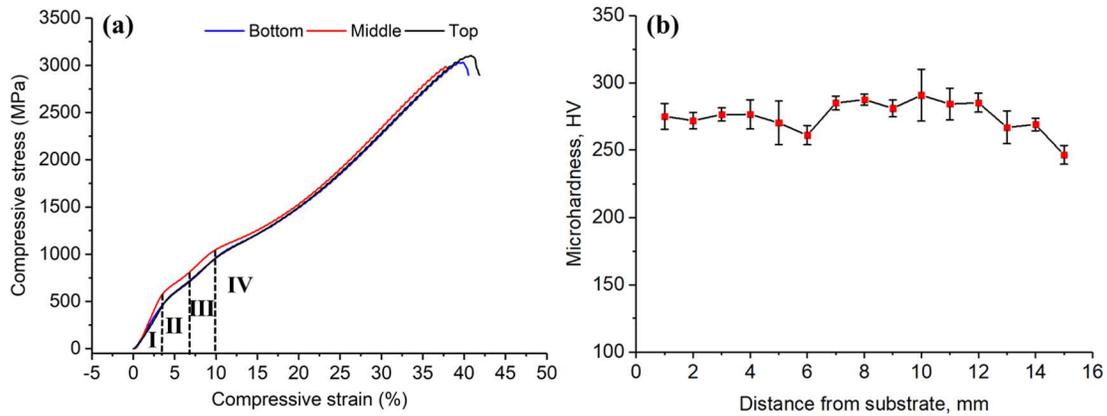

Fig. 10 Mechanical properties of: (a) engineering compressive stress-strain curves of as-deposited samples; and (b) microhardness along the scan direction (load:100 g, hold:15 s).

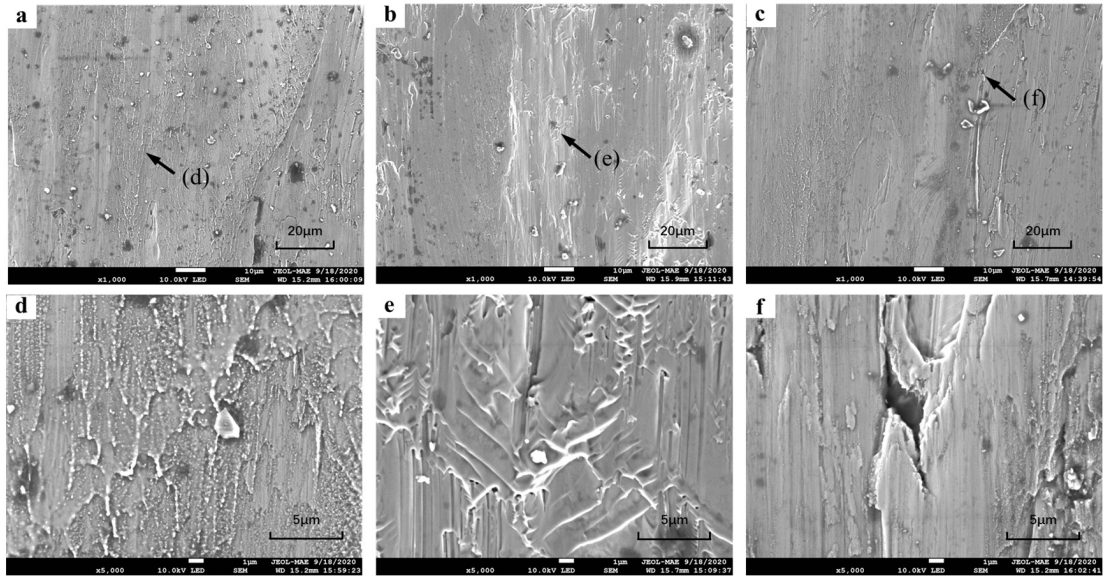

Fig. 11 SEM fractographic morphologies of compressed specimens from different regions: (a) top region, (b) middle, and (c) bottom regions; (d) dimple features; (e) cleavage features; and (f) voids.

Fig. 11 shows the SEM micrographs of the fracture surfaces for the top, middle, and bottom specimens after the compression test. As shown in Fig. 11(a) and (c), the top and bottom specimens are characterized by a smooth appearance together with plentiful dimple features. The inclusion of more dimple features corresponds to the increased ductility of the specimen, which is corroborated by the higher value of compressive strain (39.9% and 41.1%) as shown in Fig. 10(a). The middle specimen as shown in Fig. 11(b) contains fewer dimple features but with a clear cleavage facet, which indicates a less ductile nature of the sample. Hence, the middle specimen has a lower compressive strain (37.7%) than that of the top and bottom specimens. Some voids are also observed in the fabricated structure as shown in Fig. 11(f), which comes from the process defects.

## 4. Conclusions

In this study, a continuous DED method was adopted to improve the homogeneity of microstructure and thermomechanical properties for the NiTi shape memory alloy fabrication. The spatial characterization along the build direction was investigated. The FEM simulation results indicated that the different parts of the specimen experience a monotonous temperature history and nearly equal dwell time in the precipitation range. The thermal cycles (e.g., remelting or reheating peak) occurring in the traditional layer-by-layer method are eliminated. Benefiting from this characteristic, the microstructure exhibits a relatively uniform phase distribution and a similar grain morphology in the different areas. Meanwhile, the specimens from different locations demonstrate analogous transformation behaviors and similar compressive stress-strain curves with a high level of compressive strength (2982 MPa~3105 MPa ) and strain (37.7%~41.1%). The phenomenon of location-dependent thermomechanical properties usually occurred in the traditional DED method is effectively alleviated. Due to the elimination of intermittent steps and the consistency between the heat flux and solidification directions, high-quality NiTi structures with complex geometry (especially slender features) can be obtained through this continuous DED method combining multi-axis movement.

## Acknowledgments

This work was supported by the start-up fund from McCormick School of Engineering, Northwestern University, Evanston, IL, USA, and Research Grants Council of the Hong Kong Special Administrative Region, China (Project No. CUHK14202219).

## Appendix

The model parameters used in the simulation are given in Fig. A1 and Table A1:

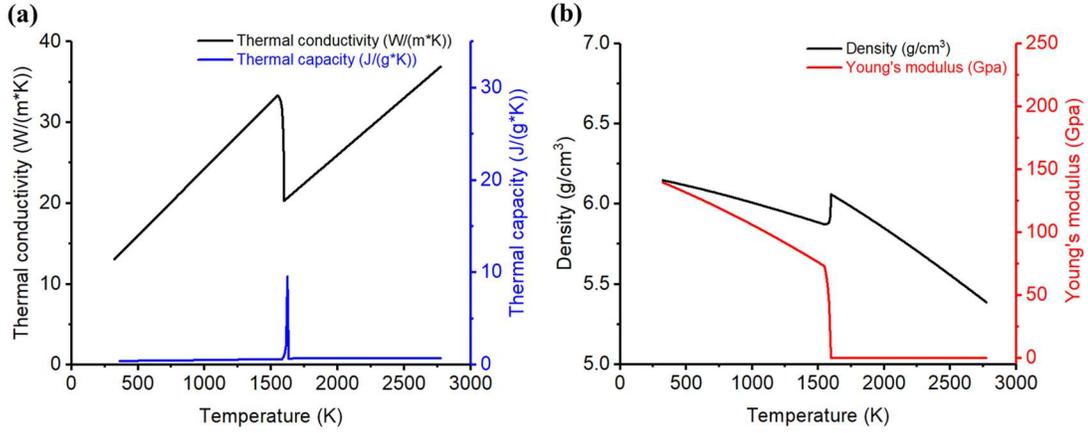

Fig. A1 Thermal-dependent material properties of NiTi alloy: (a) conductivity and capacity; and (b) density and Young's modulus.

**Table A1:** Simulation parameters

| Parameters | Values |
| --- | --- |
| Substrate width, $W$ | 50 mm |
| Substrate height, $H$ | 4 mm |
| Rod diameter, $R_{rod}$ | 1.65 mm |
| Rod height, $H_{rod}$ | 15 mm |
| Thermal expansion coefficient, $\alpha$ | 11e-6, 1/K |
| Solidus temperature, $T_s$ | 1553 K |
| Liquidus temperature, $T_l$ | 1583 K |
| Latent heat, $L_m$ | 24.2 kJ/kg |
| Power effective absorption coefficient, $\eta$ | 0.24 |
| Laser beam radius, $r_b$ | 200 μm |

The simulation parameters are partially taken from the material sheet and partially calibrated from experiments. The laser powder absorption efficiency is determined through the weight method, while the laser power effective absorption coefficient is measured through the calorimetric method. The detailed experimental calibration procedure can be referred to [58].

The temperature history presented in Fig. 3(b) was obtained by a setup using embedded thermocouples (XC-24-K-24, *OMEGA*, *USA*). A data acquisition card (USB-6001, *NI, USA*) was used to acquire the temperature signals. LabVIEW was utilized to record and post-process the temperature measurement with a sampling frequency of 10 kHz. The actual experimental setup is shown in Figure A2. The measured and simulated temperature histories were then compared to further verify the simulation model.

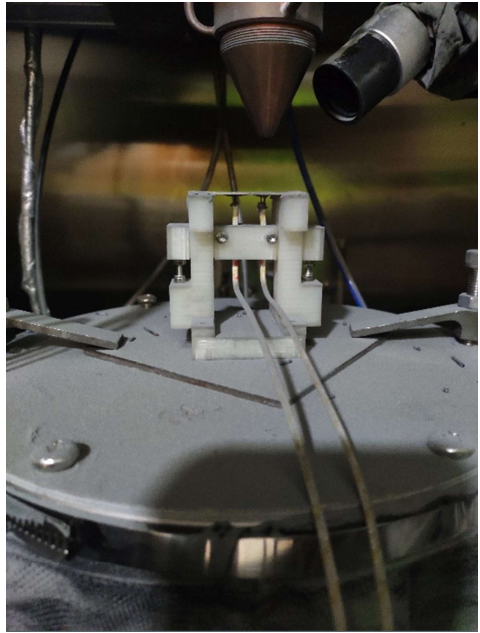

Fig. A2 Temperature measurement setup for the simulation verification.